\documentclass[11pt,a4paper]{article}
\usepackage[hyperref]{acl2021}
\usepackage{times}
\usepackage{latexsym}
\usepackage{graphicx}  
\usepackage{natbib}    
\usepackage{csquotes}  

\usepackage{microtype}

\aclfinalcopy 

\title{Detecting Interlocutor Confusion in Situated Human-Avatar Dialogue: A Pilot Study}

\author{
Na Li, John D. Kelleher, Robert Ross\\
  School of Computing\\
  Technological University Dublin  \\ 
  \{na.li, john.d.kelleher, robert.ross \}@tudublin.ie \\
 }

\date{}

\begin{document}
\maketitle
\begin{abstract}
In order to enhance levels of engagement with conversational systems, our long term research goal seeks to monitor the \emph{confusion state} of a user and adapt dialogue policies in response to such user confusion states. To this end, in this paper, we present our initial research centred on a user-avatar dialogue scenario that we have developed to study the manifestation of confusion and in the long term its mitigation. We present a new definition of confusion that is particularly tailored to the requirements of intelligent conversational system development for task-oriented dialogue. We also present the details of our Wizard-of-Oz based data collection scenario wherein users interacted with a conversational avatar and were presented with stimuli that were in some cases designed to invoke a confused state in the user. Post study analysis of this data is also presented. Here, three pre-trained deep learning models were deployed to estimate base emotion, head pose and eye gaze. Despite a small pilot study group, our analysis demonstrates a significant relationship between these indicators and confusion states. We understand this as a useful step forward in the automated analysis of the pragmatics of dialogue. 
\end{abstract}

\section*{Keywords}
\textbf{Confusion detection, situated dialogues, emotion recognition, head pose, eye gaze, pragmatics, avatar, wizard-of-oz}

\section{Introduction}
\label{sect:intr}

Situated conversation either in the case of Human-Robot Interaction (HRI) or in the virtual world with Avatars provides significant challenges and opportunities for the further development and deployment of dialogue systems. In the case of robotic systems, applications ranging from healthcare assistants \cite{healcarehumanrobot2020} to tour guides in a museum \cite{tourguide2019} can all take advantage of spoken interaction in a situated setting. Meanwhile within an online setting, applications such as online learning system \cite{engagementHCI2019} can also take advantage of the speech channel. However, in each of these scenarios, the need for fluid interaction where users remain engaged is hugely important, and that in order to provide an effective interface, the dialogue system must respond appropriately to the user's words and mental states.

Confusion is a unique mental state that can either precede a high degree of positive engagement in a task, or can also be correlated with negative states such as boredom and subsequent disengagement from a conversation \cite{DMelloConfusionlearning2014}. Estimating the confusion state of a user can hence be a very important step in improving the pragmatics modelling properties of an interactive system. By checking for confusion, or indeed precursors of confusion, we can in principle adjust the dialogue policy or information being presented to the user in order to assist them in the specific task being undertaken. Such monitoring can be seen as a specific form of engagement detection \cite{engagementConnect,Dewan2018EngagementDI}. In mainstream Human-Computer Interaction (HCI) studies, there have to this point been a number of studies that have investigated the modelling and detection of confusion \cite{eegconfusionlevel,Grafsgaard2011,Zhou2019}. However, the majority of study in this area has concerned online learning in specific environments such as AutoTutor, ITS (Intelligent Tutoring Systems) and MOOCs (Massive Open Online Courses) or serious games; little work has focused on general engagement or task-oriented dialogue.

In light of the above, our goal in this paper is to explore the potential manifestations of confusion and investigate whether it is possible to detect the confusion state as a pragmatics analysis task to supplement a multimodal situated dialogue system. While our primary area of interest is in HRI, this study has been executed with a focus on Human-Avatar Interaction (HAI). This is in part due to the relative ease of executing avatar based studies without the physical robot. More specifically, two research questions in this study are presented:
\begin{enumerate}
    \item Are participants aware they are confused if we give them a specific confusing situation?
    \item Do participants express different physical or verbal/non-verbal behaviours when they are confused that we can detect?
\end{enumerate} 

To answer these research questions, a wizard-of-oz human-avatar interaction study was designed based around an avatar web application which allowed participants to be both recruited remotely and engage in the interaction remotely. Study stimuli included a series of situated conversations to attempt to trigger confused states in the participants. Participants' behaviours including verbal or non-verbal languages, facial expression and body pose were recorded and subsequently analysed. Before detailing this work, we begin with a review of related work with a particular focus on setting out a relevant framework for engagement and specifically confusion estimation. 

\section{Related Work}
\label{sect:litreview}

The detection and monitoring of a participant's mental state in conversational interaction is a well-established area of research. In this section, we briefly review a number of works related to our own area of focus, and look in particular at the challenge of defining and identifying confused states during interaction. 

\subsection{Emotion \& Engagement Recognition}
\label{ssec:emotionRec}

The recognition of human emotional states has been noted as a pillar in engaging conversation in domains such as human-robot interaction \citep{emotionrecognitionHRI2020}. In early work, \citet{FoundhumancomputingFacialEmotion2007} indicated that human emotion may not be directly observable because emotion is a cognitive state, but that emotion can be explained through interaction context and evidenced by user survey, behavioural and physiological indicators. In terms of physiological indicators of emotional state, the facial expression is the most natural  manifestation for a human. In terms of analysing facial expressions, the Facial Action Coding System (FACS) \cite{FoundhumancomputingFacialEmotion2007, FACS2017} is an example of a \emph{part-based method}, which defined the smallest units of muscular activity that can be observed in the human face, called Action Units (AUs). FACS is a well-known analysis tool that has been combined with self-report measurements. More recently of course, Deep Learning based image analysis methods have been used to demonstrate high accuracy for emotion recognition on facial images \cite{ emotionfacailexpression2019}. Similarly various recurrent and ensemble network architectures have been built to analyse multimodal datasets including speech (audio) data, text-based data and video data to provide estimates of emotional state \cite{Iemocapemotion2018, dialogueCNNemotion2018}. 

Beyond facial or verbal expressions, certain behaviours such as head pose, and eye gaze are also noted as non-verbal indicators of engagement and mental state during interaction. In particular, \citet{eyegazingEMERY2000581} explained that eye gaze is a component of facial expressions that can be interpreted as a cue to show people's attention to another individual, events or objects -- either within spoken interaction or other domains \cite{adaseyegazing,verbalnonverbalHRI2015,Zhang2020ETHXGaze}. Similarly, head-pose estimation is also studied extensively in face-related vision research where it is considered related to vision and eye-gaze estimation. \citet{headposesurvey2009} provided an example of Wollaston illusion, where although eyes are in the same directions, the eye-gazing direction is decided by two differently oriented heads. They indicated that people with different head poses can reflect more emotional information such as dissent, confusion, consideration and agreement. Meanwhile, methods for training models of eye-gaze and head-pose estimation are generally consistent with facial expression analysis. 

\subsection{Engagement Detection}
\label{ssec:engageDetect}

For us, confusion detection is intended to enhance engagement, and engagement in interaction is widely studied within the fields of psychology and linguistics, where engagement, for example, can be recognized as being broken down into three aspects: social connection  \citep{engagementHCI2019,engagementConnect}, mental state centric \citep{engagementConnect}, and a motivated or captivated phenomena \citep{engagementdefine_media}. For our purposes here, a key challenge is the detection of engagement. Within HCI and HRI there are three basic categories of engagement detection which are manual, semi-automatic and automatic \citep{Dewan2018EngagementDI}. Manual detection usually refers to tasks such as participant self-reporting or observational check-lists. Semi-automated engagement monitoring utilizes the timing and accuracy of responses such as reaction time for an interaction, or judgements of users' responses to tasks within an interaction. The automatic category meanwhile typically refers to machine learning driven methods operating directly or raw data or automatically extracted features, \emph{e.g.}, \citet{onflydetection}. 

In recent years, there have been a wide variety of studies that have attempted to estimate and make use of user engagement in interaction \cite{childengagement}. For example, \citet{uehri2017} studied a human-robot interaction scenario with a fully automated robot (\emph{i.e.}, Pepper Robot) for recognizing users' engagement in spontaneous conversations between users and the robot.
 
\subsection{Confusion Detection}
\label{ssec:confusionDetect}

As a psychological state, confusion has been studied mostly to date within the context of pedagogy and related applied fields of learning, and depending on context has been defined as everything from a bonafide emotion through to an epistemological state. When confusion is considered as an effective response, confusion happens in people who are enthusiastic to know or understand something \citep{DMello2014}. On the other hand, confusion may be defined as an epistemic emotion \cite{Lodge2018} that is associated with blockages or impasses in the learning process. Confusion is also triggered by cognitive disequilibrium, where cognitive disequilibrium is itself defined as a state wherein a participant is learning but where obstacles to the normal flow of the learning process are encountered, the participant may feel confused when they encounter contradictory information leading to uncertainties, resulting in cognitive disequilibrium \cite{YangMOOCs}. 

\citet{Arguel2015} presented two thresholds ($T\_a$ and $T\_b$) for levels of confusion in learning. The level of confusion between the two thresholds is termed productive confusion. It indicates that learners are engaged in overcoming their confused state. However, when the level of confusion is over $T\_b$ (persistent confusion), it is easy for learners to move to a state of frustration or even boredom. If the level of confusion is less than $T\_a$, then learners may continue to engage in their learning. \citet{Lodge2018} designed a learning event in which the learner was in cognitive disequilibrium, where the disequilibrium was created by a manufactured impasse in the learning process. Similar to the notion of thresholds, in this study learners could be categorised to be in the zone of optimal confusion or sub-optimal confusion. Optimal confusion is productive confusion, which indicates that the learners are engaged in overcoming their confused state. On the other hand sub-optimal confusion is associated with persistent confusion where learners could not resolve the disequilibrium which in turn leads to possible frustration or boredom. \citet{DMello2014} meanwhile offers a transition oriented model where confusion can be seen as part of emotional transition between engagement/flow and frustration/boredom. 

While there have been a number of studies that have touched on confusion in these learning scenarios, we find that there is no well-documented definition of confusion that can assist this research in modelling and mitigating confusion in interaction. In light of this, and for use in the context of dialogue centric human-machine interaction, we offer the following working definition of confusion. \emph{Confusion is a mental state where under certain circumstances, a human experiences obstacles in the flow of interaction. A series of behaviour responses (which may be nonverbal, verbal, and, or non-linguistic vocal expression) may be triggered, and the human who is confused will typically want to solve the state of cognitive disequilibrium in a reasonable duration. However, if the confusion state is maintained over a longer duration, the interlocutor may become frustrated, or even drop out of the ongoing interaction.}

While in an ideal interaction there would be little or no confusion in practice, for the purpose of study it is useful to be able to induce confusion in an interlocutor. Within the literature, at least four types of confusion induction have been considered \cite{LEHMAN2012184, Silvia2010ConfusionAI}. The first is \emph{complex information} where the material to be understood is genuinely complex and presents challenges to one individual (that might not necessarily apply to another individual). The second is the challenge of \emph{contradictory information} where inconsistencies may push an individual into a confused state. The third case is the provision of \emph{insufficient information} where confusion is due simply to not enough background or relevant information being provided to an individual. Finally, and related to contradictory information, we have \emph{feedback} inconsistencies where during an interaction one agent provides another with information that is inconsistent with the interaction to date. 

\section{Study Design}

With our working definition of confusion as a guideline, we designed a Wizard of Oz (WoZ) \citep{Riek2012WizardOO} study to investigate: (a) the effectiveness of confusion induction methods in interactions; and (b) the relative performance of a range of manual, semi-automatic and automatic methods for confusion estimation. In the following we describe our overall experiment design; stimuli design, and approach to data analysis. 

\subsection{Study Overview}
\label{sect:experimentDesign}
\label{ssec:method}

While our main focus is in the context of human-robot interaction, this experiment was designed as a human-avatar study to account for some of the study design limitations that were experienced due to the COVID-19 pandemic of 2020-2021. While an avatar does not provide a full situated experience, an avatar has notable benefits over a speech or text only based interaction \cite{avatarDialogue2017}. 

The experiment was based on a semi-spontaneous one-to-one conversation between an agent (in our case a wizard controlled avatar) and a participant (the user). Participants were recruited from across the university and study programmes, and these participants remained in their own homes while the wizard was similarly in their own work environment. Participants were requested to connect via a camera and audio enabled laptop and with reliable internet connectivity. Typical participation times were designed to be less than 15 minutes in total with 5 minutes for the task centric part of the conversation. At the beginning of the interaction participants were given consistent instructions and consent forms, and following the task (described later) all participants were asked to complete a survey (also detailed later). Finally at the end of this experiment, each participant was invited for a 3-minutes interview with the researcher.

\begin{figure}
  \centering
  \includegraphics[width=0.43\textwidth,height=3.5cm]{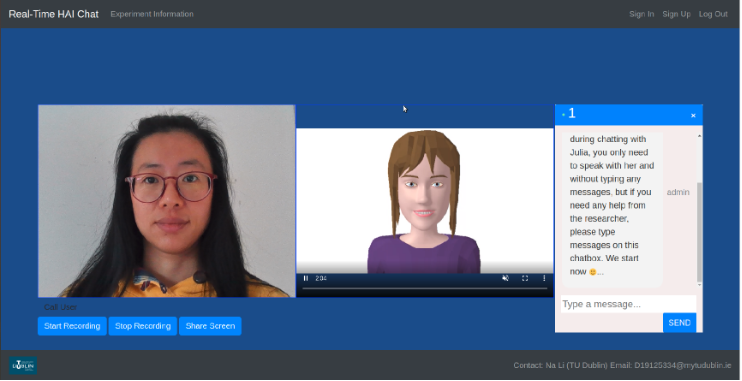}
  \caption{HAI Real-Time Chat Web App}
  \label{fig:webapp}
\end{figure}

For this study, a web application framework was developed and built on two components: one was a real-time chat application, while the other was an avatar application that was embedded within the real-time chat application. The avatar application was based on the framework by  \citet{Sloan2020Emotional} which provides a sandbox with modules of an e-learning platform with an animated avatar. It integrates animation, speech recognition and synthesis, along with full control of the avatar's facial expressions to convey happiness, sadness, surprise, etc. The real-time chat application meanwhile is a web application for online interaction between the agent/avatar and participant that we developed to handle all survey steps, communication, and enrolment with the user. The application is designed to enable full data recording of both the avatar and the remote participant's audio, text, and camera streams. We illustrate the complete framework in Figure \ref{fig:webapp}. 

\begin{table}
\centering
\begin{tabular}{lrl}
\hline \textbf{Participant 1}  \\ \hline
\hline \textbf{Stimulus} & \textbf{Task} & \textbf{Condition} \\ \hline
1st & Task 1 & A \\
2nd & Task 2 & B \\
3rd & Task 3 & A \\
\hline
\hline \textbf{Participant 2}  \\ \hline
\hline \textbf{Stimulus} & \textbf{Task} & \textbf{Condition} \\ \hline
1st & Task 1 & B \\
2nd & Task 2 & A \\
3rd & Task 3 & B \\
\hline
\end{tabular}
\caption{\label{tab:experimentsequ} An example of the experiment sequence for two separate study participants.}
\end{table}

There were 23 participants in six countries who participated in this study; three of the participants were unable to complete the final experiment due to internet connectivity or equipment problems. All participants were over 18 years of age from different colleges around the world who can have a simple conversion in English at least. We successfully collected video data, user surveys and demographics information from 19 participants (8 males, 11 females) and acquired their permission to use their data for this research purpose.

\subsection{Dialogue Design}
\label{ssec:dialogue}

To stimulate confusion within a short conversation, we defined two conditions with appropriate stimuli. In condition A, stimuli were designed to invoke confusion in the participant. In condition B, stimuli were designed to allow a participant to complete a similar task in a straightforward way and should avoid confused states. Three separate task sets were defined with each task designed for both conditions. Task 1 was a simple logical question; task 2 was a word problem; while task 3 was a math question. We prepared at least two questions for each conditions in each task. As for the sequence of the experiment, Table \ref{tab:experimentsequ} shows the sequence of conditions for each participant; for the first participant for example, the sequence of conditions is Task 1 with condition A, Task 2 with condition B and task 3 with condition A. The sequence of conditions between participants was alternated to balance the number of conditions for data analysis.

As for situated dialogues, there are two dialogues corresponding to two conditions, and one dialogue is for one condition; three patterns of confusion for two conditions: the first pattern of complex information and simple information, the second pattern of insufficient information and sufficient information and the last pattern of of correct-negative feedback and correct-positive feedback. For example, below is a word problem with the second pattern, for insufficient information in condition A: \emph{``There are 66 people in the playground including 28 girls, boys and teachers. How many teachers were there in total?''}; while the case for sufficient information \emph{i.e.}, condition B is: \emph{``There are 5 groups of 4 students, how many students are there in the class?''}.

\begin{figure}
  \includegraphics[width=0.45\textwidth,height=6cm]{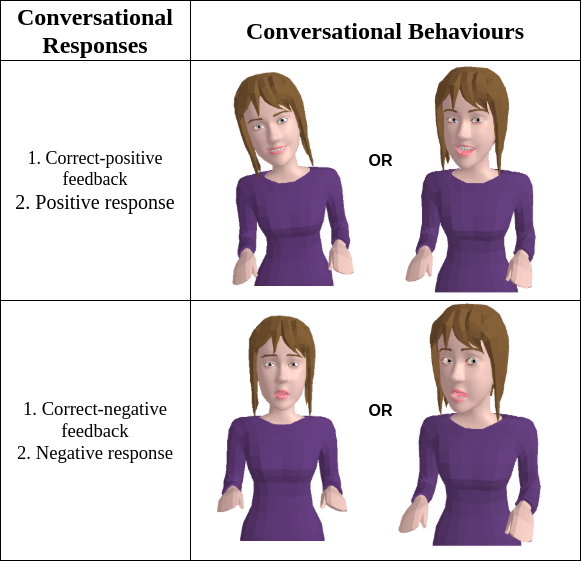}
  \caption{The mapping of the reaction status and visible traits for the avatar.}
  \label{fig:mappingresponse}
\end{figure}

It should be noted that the design of individual stimuli includes both the verbal and non-verbal elements of the interaction. Thus, avatar's responses were mapped to visible behaviours \cite{Cassell2004FullyEC}. Figure \ref{fig:mappingresponse} shows an example of the mappings of the avatar's facial expressions and body gestures for conversational responses and conversational behaviours corresponding to positive reaction and negative reaction.

\subsection{Data Preparation}
\label{ssec:dataprocess}

\begin{figure}
  \centering
  \includegraphics[width=0.48\textwidth,height=3.5cm]{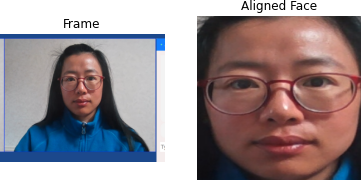}
   \caption{Frame and aligned facial image}
   \label{fig:alignimage}
\end{figure}

Frame data was extracted for 19 participants' videos and each video was labelled with the sequence of conditions (\emph{e.g.}, ABA or BAB), such that all frames were labelled as either condition A or condition B. To verify the frame labelling, labelling files with frame names and one condition were manually matched. The image data for condition A had 4084 frames, while the image data for condition B has 3273 frames. Facial recognition and alignment are a significant first step in pre-processing frame data, thus we applied an efficient method to centre crop a 224x224 region for each frame \cite{facialemotion2021}, and then used a (Multi-task Cascaded Convolutional Neural Networks) MTCNN-based face detection algorithm without frame margins. Figure \ref{fig:alignimage} shows the original frame on the left, with the processed image on the right. 

In addition to making use of the raw frame data, we also involved use of the post interaction survey questions. Here the user survey consisted of 10 questions using a 5-level Likert scale. Three of the questions were specific to the three tasks (logical questions, word problems and math questions) including the scores of the both conditions. Each user survey contains the results of the two conditions. Thus, the results of the survey were separated into two independent groups by the two conditions and then collected into one file for analysis with a flag noting ``condition'', as well as additional parameters such as the average of scores of two tasks under the same condition.


\section{Data Analysis}
\label{ssec:analysis}

To address our research questions introduced earlier, we applied a number of feature analysis algorithms to our data and analysed the interactions between these and our experimental conditions and the results of the survey questions. Below we detail these methods and present the results of this analysis. 

\subsection{Frame Data Measurement}

\begin{figure}
  \centering
  \includegraphics[width=0.4\textwidth,height=3.8cm]{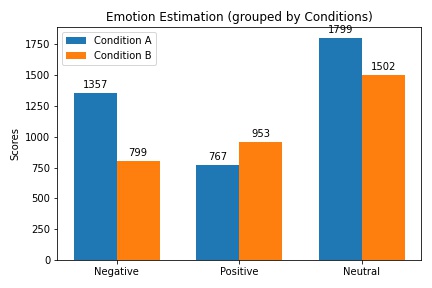}
   \caption{Comparison of three emotional categories grouped by condition A and condition B}
   \label{fig:emotion_3}
\end{figure}

Our primary form of analysis was based around the automated processing of video data to determine if automatically extracted indicators of emotion, head pose and eye gaze have a significant correlation with confusion state. 
For emotion detection we made use of a visual emotion detection algorithm \cite{facialemotion2021} based on the MobileNet \cite{MobileNet2017} architecture and trained on the AffectNet dataset \cite{AffectNet2017} for 8 target classes, namely the 7 main facial expressions: Neutral, Happy, Sad, Surprise, Fear, Anger, Disgust, and an 8th: Contempt. Table \ref{tab: emtion_7} shows the number of each of the 7 primary emotion categories predicted grouped by condition A and condition B. It is notable that the predicted results in condition A for 4 negative emotion categories (anger, disgust, fear and sadness) are considerably more than for condition B. In contrast, as for 2 positive emotion categories  (happiness and surprise), the number of predicted results in condition A are lower than condition B. Undoubtedly, for neutral emotion we see that the condition A is higher than condition B. Figure \ref{fig:emotion_3} presents a comparison of the results of emotion prediction for three categories (negative, positive and neutral) grouped by the conditions. In order to deep understand the correlation relationship between the three emotional categories and conditions, a statistical analysis of whether there is a statistically significant relationship between three emotional categories and the two experimental condition classes A and B. The result of an independent-sample t-test is that there is a significant difference in the three emotional categories (negative, positive and neutral) and two conditions ($M=0.77, SD=0.94$ for condition A, $M=0.48, SD=0.60$ for condition B), $t(715)=5.05, \rho-value<0.05$.
\begin{table*}
    \begin{tabular}{l|l|l|l|l|l|l|l|l}
    \hline \textbf{Condition} & \textbf{Anger} & \textbf{Disgust}  & \textbf{Fear} & \textbf{Sadness} & \textbf{Happiness}  & \textbf{Surprise}& \textbf {Neutral} & \textbf{Overall}\\ \hline
      A  &  262 & 282 & 136 & 677 & 702 & 65 & 1799 & 3923\\
      B  &  77 & 165 & 57 & 480 & 858 & 95  & 1502 & 3234 \\
      \hline
    \end{tabular}
    \caption{Result of emotion estimation grouped by condition A and condition B}
    \label{tab: emtion_7}
\end{table*}

For head-pose estimation, we applied use of the model due to  \citet{headpose2017} that makes use of CNNs, dropout and adaptive gradient methods trained on the three public datasets, namely the Prima head-pose dataset \cite{primaheadposedataset2004}, the Annotated Facial Landmarks in the Wild (AFLW) \cite{aflwdataset2011} and the Annotated face in the Wild (AFW) dataset \cite{afwdataset2012}. The predicted results are the angles of pitch, yaw and roll for each image. We calculated the sum of absolute values of the three angles as a new feature for statistical analysis because only sum of values of pitch, yaw and roll will be the canceling effect of the positive and negative values, even the sum of values may be 0 as a person has different angles of direction with different positive or negative values of pitch, yaw and roll. Using this metric our related research question is whether there is a statistically significant relationship between the sum of absolute values of these three angles and our two experimental condition classes A and B. The result of an independent-sample t-test is that there is a significant difference in the sum of absolute values of these three angles and two conditions ($M=21.96, SD=9.46$ for condition A, $M=27.40, SD=12.21$ for condition B), $t(703)=-6.61, \rho-value<0.05$.

\begin{figure}
  \centering
  \includegraphics[width=0.5\textwidth,height=4cm]{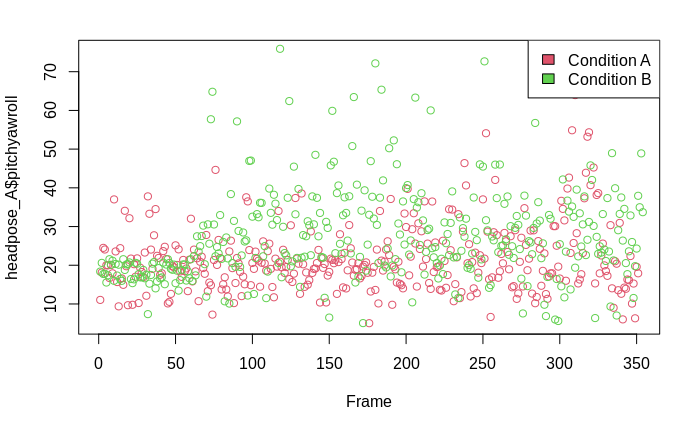}
   \caption{Head-pose estimation: plot the sum of angles values for condition A and condition B }
   \label{fig:headposeplot}
\end{figure}

\begin{figure}
  \centering
  \includegraphics[width=0.5\textwidth,height=3.8cm]{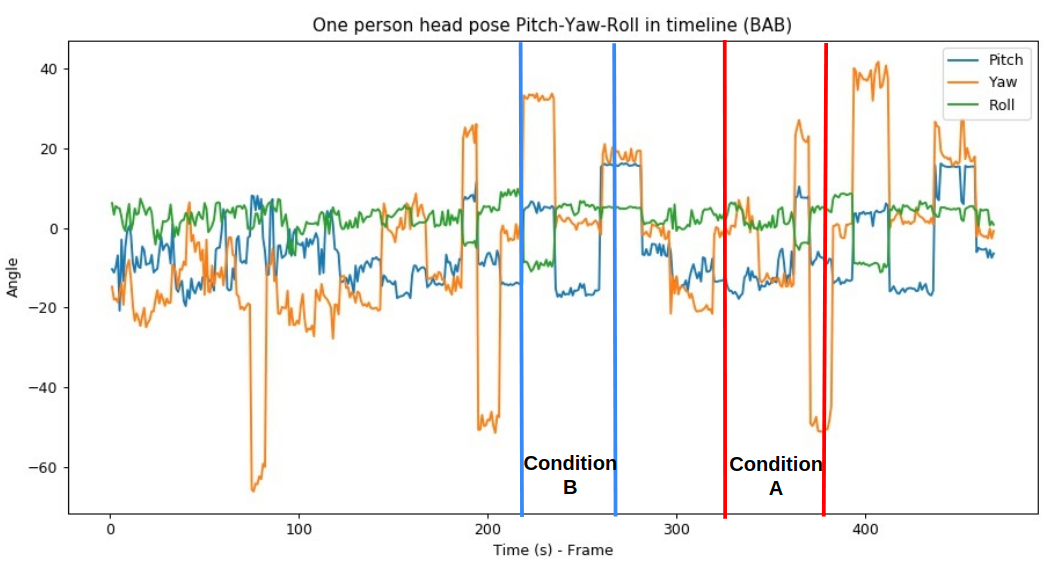}
   \caption{Head-pose estimation: plot the change of one person's pitch, yaw and roll angles at an experiment}
   \label{fig:headposeplot_persion}
\end{figure}

We also plotted the sum of the three angles of the two conditions (see Figure \ref{fig:headposeplot}). From this we can see that the values of condition A form a less discrete distribution than condition B. While we cannot draw conclusions from it, we also analysis the specific yaw, roll and pitch angle for individuals. To illustrate Figure \ref{fig:headposeplot_persion} shows the labelled time for condition A (read lines) and condition B (blue lines), thus this shows for one individual the fluctuations of the pitch angle, yaw angle and roll angle in the time series. This indicates that the angle of the participant's head posture angle in condition A was generally smaller than the angle of the participant's head posture angle in condition B.

For eye-gaze estimation we applied a state-of-art gaze estimation model that had been trained on the recently published ETH-XGaze dataset \cite{Zhang2020ETHXGaze}. The GTH-XGaze dataset includes more than one million high-resolution images of different gazes in extreme head poses from 11 participants. The predicted results are angles of pitch and yaw for relative eyes directions. Again in this case, we summed up the absolute angles of two results, and we ask whether there is a relationship between this metric and our two experimental conditions A and B. An independent-samples t-test again was conducted to compare the two sets. There is a significant difference in the sum of absolute values of pitch and yaw and two conditions was found ($M=0.44, SD=0.26$ for condition A, $M=0.49, SD=0.22$ for condition B), $t(728)=-2.58, \rho-value<0.05$. 

\begin{figure}
  \centering
  \includegraphics[width=0.5\textwidth,height=4cm]{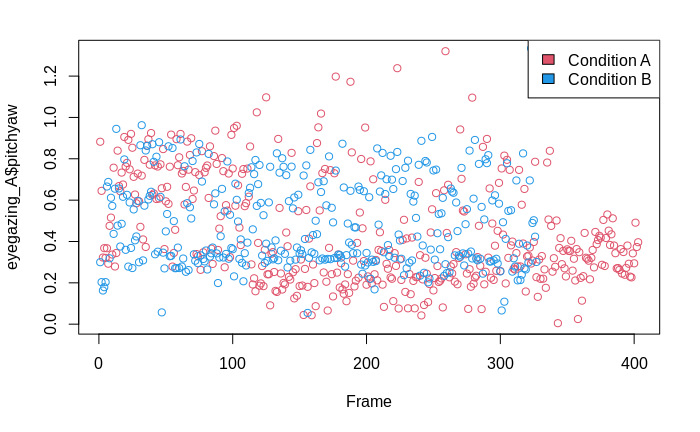}
   \caption{Eye-gaze estimation: plot the sum of angles values for condition A and condition B }
   \label{fig:eyegazeplot}
\end{figure}

\begin{figure}
  \centering
  \includegraphics[width=0.5\textwidth,height=3.8cm]{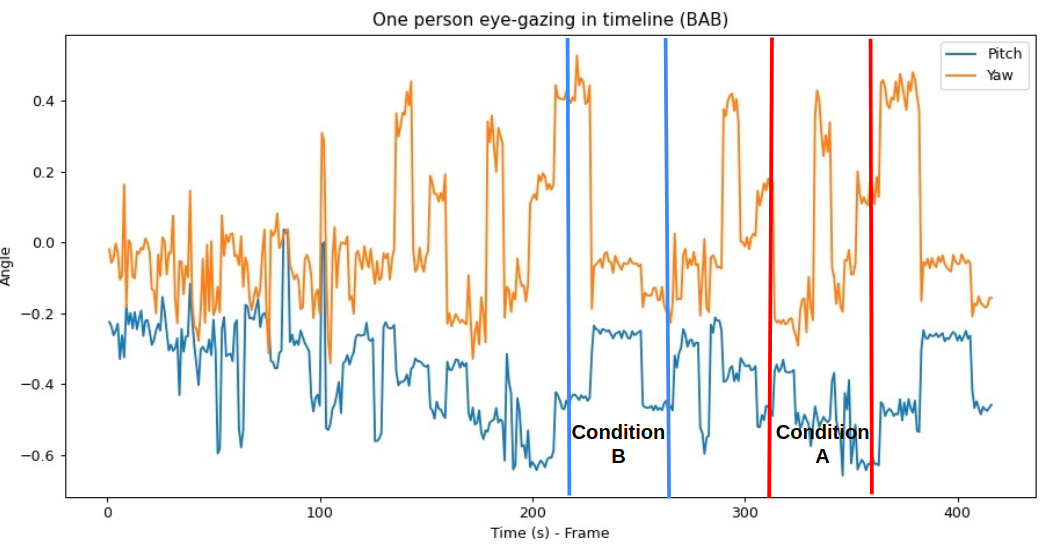}
   \caption{Eye-gaze estimation: plot the change of one person's pitch and yaw angles
   at an experiment}
   \label{fig:eyegazelot_persion}
\end{figure}

In addition, we used the same method as with head-pose estimation to demonstrate these results. Figure \ref{fig:eyegazeplot} shows that the eye-gaze values of condition A form a more discrete distribution than those for condition B. Meanwhile in figure \ref{fig:eyegazelot_persion}, we can see that the fluctuations of the same individual participant’s pitch angle and yaw angle plotted in the time series. Here we see in this example case that the gaze angle of the participant in condition A is greater than that of the participant in condition B. 

\subsection{Subjective Measurement}
For our survey results we analysed self-reported scores with respect to the two stimuli control conditions A and B. We can break down this analysis into two sub-questions. The first of these is whether there is a statistically significant relationship between the average of self-reported confusion scores for each of the three performed tasks and the conditions. The second three sub-questions are whether there is a statistically significant relationship between confusion scores for each of the three performed tasks and the two conditions. 

With respect to the first question, an independent-samples t-test was used and found that there is no significant difference between the average of confusion scores of the three tasks and two conditions ($M=3.50, SD=1.40$ for condition A, $M=2.97, SD=1.12$ for condition B), $t(36)=1.28$, $\rho-value=0.21$. However, with respect to the second three questions: firstly, there is no significant difference in the confusion scores for task 1 with two conditions was found ($M=3.00, SD=1.07$ for condition A, $M=2.44, SD=1.33$ for condition B), $t(15)=0.94, \rho-value=0.36$; secondly, there is no significant difference in the confusion scores for task 2 with two conditions was found ($M=3.09, SD=1.22$ for condition A, $M=3.10, SD=1.29$ for condition B), $t(19)=-0.02, \rho-value=0.99$; lastly, the result indicated that there is a significant difference in the confusion scores for task 3 was found ($M=4.38, SD=0.74$ for condition A, $M=3.00$, $SD=1.12$ for condition B), $t(15)=2.94$, $\rho-value<0.05$.

\section {Discussion}
\label{sect:discussion}

Based on the results provided in the previous section, we note that the following holds with respect to the specific questions that we identified:
\newline 1. Participants are not always aware they are confused if we gave them a specific confusing situation.
\newline 2. When they are confused, their emotion is more negative than when they are not confused.
\newline 3. When they are confused, the range of angles of eye gazing is more than when they are not confused.
\newline 4. When they are confused, the range of the angles of head shaking is less than when they are not confused.
 

Due to size and scope limitations, this is in essence a pilot study of confusion induction and detection. Notable limitations are not only on sample size but a number of technical challenges with the study. First, the qualities of videos of participants varied because of the quality of network connection, camera specification, and camera position, etc. Second, the sample size and range of participant backgrounds are a major limitation which limits the conclusions that can be drawn from this work. Third, as noted in Section \ref{ssec:confusionDetect}, confusion is a unique mental state which can transit to positive states or negative states; in this pilot study, there are no clear dialogues' boundaries and time frames to distinguish the level of confusion. Finally, it should be mentioned that during the 3-minutes interview, many participants reported that they expected to have a wonderful conversation with the avatar, but this experiment lacked casual conversation and even a participant's expectations of the avatar's abilities were often not met.

Nevertheless, we believe that the study results do demonstrate that the approach to data collection and analysis are worthwhile, moreover we intend to build upon this in future work. At a minimum we intend to introduce audio and linguistic content analysis to expand beyond the visual and self-reporting data made use of in the current work. Second, and importantly, having established the general framework we wish to conduct further in-person studies to build upon our framework but with fewer constraints in place due to the COVID-19 pandemic. Ultimately our goal is also to study mitigation factions in confusion situations, and as such we will also be expanding our studies to study the effects of different clarification strategies on the confusion state. 

\section{Conclusion}
\label{sect:conclusion}

In this paper we have proposed the study, detection, and mitigation of confusion as an important factor in improving dialogue centric human computer interaction. We also proposed a new working definition of confusion for our purposes and outlined a study that we conducted to determine if confusion could be induced and detected in a human-avatar task oriented interaction. While we did not find a significant relationship between self-reported confusion scores and induced confusion states, we did find significant differences between observed physical states, \emph{i.e.}, facial emotion, head pose, eye gaze and our induced confused states. Although a small sample size is insufficient for generalisation, we see this work as a crucial initial step down the path to a computational model of confusion in multimodal dialogue. 

\section*{Acknowledgments}
This publication has emanated from research conducted with the financial support of Science Foundation Ireland under Grant number 18/CRT/6183. For the purpose of Open Access, the author has applied a CC BY public copyright licence to any Author Accepted Manuscript version arising from this submission.
\bibliographystyle{acl_natbib}
\bibliography{acl2021}

\end{document}